\documentclass{ws-procs9x6}

\newcommand{\ba}{\begin{eqnarray}}
\newcommand{\ea}{\end{eqnarray}}
\newcommand{\bmath}{\begin{subequations}}
\newcommand{\emath}{\end{subequations}}
\newcommand{\ban}{\begin{eqnarray*}}
\newcommand{\ean}{\end{eqnarray*}}
\newcommand{\tl}{\tilde{\ell}}
\begin{document}

\title{
Relativistic Pseudospin Symmetry as\\ 
a Supersymmetric Pattern in Nuclei}

\author{A. Leviatan}

\address{
Racah Institute of Physics, The Hebrew University, Jerusalem 91904, Israel\\
E-mail: ami@phys.huji.ac.il}

\maketitle

\abstracts{
Shell-model states involving several pseudospin doublets and 
``intruder'' levels in nuclei, are combined into larger multiplets. 
The corresponding single-particle spectrum exhibits a supersymmetric 
pattern whose origin can be traced to the relativistic pseudospin symmetry 
of a nuclear mean-field Dirac Hamiltonian with scalar and vector 
potentials.} 

\section{Introduction}
Pseudospin doublets~\cite{arima69} 
in nuclei refer to the empirical observation 
of quasi-degenerate pairs of certain 
shell-model orbitals with non-relativistic 
single-nucleon radial, orbital, and total angular 
momentum quantum numbers:
\ba
(n,\ell,j = \ell + 1/2) 
\quad {\rm and} \quad
(n-1,\ell + 2,j^{\prime} &=& \ell + 3/2) ~.
\label{psdoub}
\ea
The doublet structure (for $n\geq 1$) 
is expressed in terms of a ``pseudo'' orbital
angular momentum $\tl = \ell + 1$ and ``pseudo'' spin, 
$\tilde {s} = 1/2$, coupled to $j = \tl - 1/2$ and $j^{\prime}=\tl+1/2$. 
For example, $[n s_{1/2},(n-1)\,d_{3/2}]$ will have $\tl= 1$, etc. 
The states in Eq.~(\ref{psdoub}) involve only normal-parity shell-model 
orbitals. The states $(n=0,\ell,j=\ell+1/2)$, 
with aligned spin and no nodes, 
are not part of a doublet. 
This is empirically evident in heavy nuclei, where such states 
with large $j$, {\it i.e.}, $0g_{9/2},\;0h_{11/2},\;0i_{13/2}$, are the 
``intruder'' abnormal-parity states, which are unique in the major shell. 

Pseudospin symmetry is experimentally well corroborated and plays a 
central role in explaining features of nuclei~\cite{bohr82} including 
superdeformation~\cite{dudek87} and identical bands~\cite{naza90}. 
It has been recently shown to result from a
relativistic symmetry of the Dirac Hamiltonian in which the sum of the 
scalar and vector nuclear mean field potentials cancel~\cite{gino97}. 
The symmetry generators~\cite{ginolev98} 
combined with known properties of Dirac bound states, provide  
a natural explanation~\cite{levgino01} 
for the structure of pseudospin doublets 
and for the special status of ``intruder'' levels in nuclei.  

Figure 1 portrays the level scheme of an ensemble of pseudospin 
doublets, Eq.~(\ref{psdoub}), with fixed $\ell$, $j$, $j^{\prime}$ 
and $n=1,2,3,\ldots$ together with the ``intruder" level 
$(n=0,\ell,j=\ell+1/2)$. 
The single-particle spectrum exhibits 
towers of pair-wise degenerate states, 
sharing a common $\tilde{\ell}$, and an additional non-degenerate 
nodeless ``intruder'' 
state at the bottom of the spin-aligned tower.
A comparison with Fig.~2 reveals a striking similarity with 
a supersymmetric pattern. 
In the present contribution we identify~\cite{lev04} 
the underlying supersymmetric structure associated with a Dirac 
Hamiltonian possessing a relativistic pseudospin symmetry.  
\noindent
\begin{figure}[t]
\begin{minipage}{0.49\linewidth}
\vspace{-0.35cm}
\epsfxsize=3.9cm
\begin{rotate}{-90}
\epsfbox{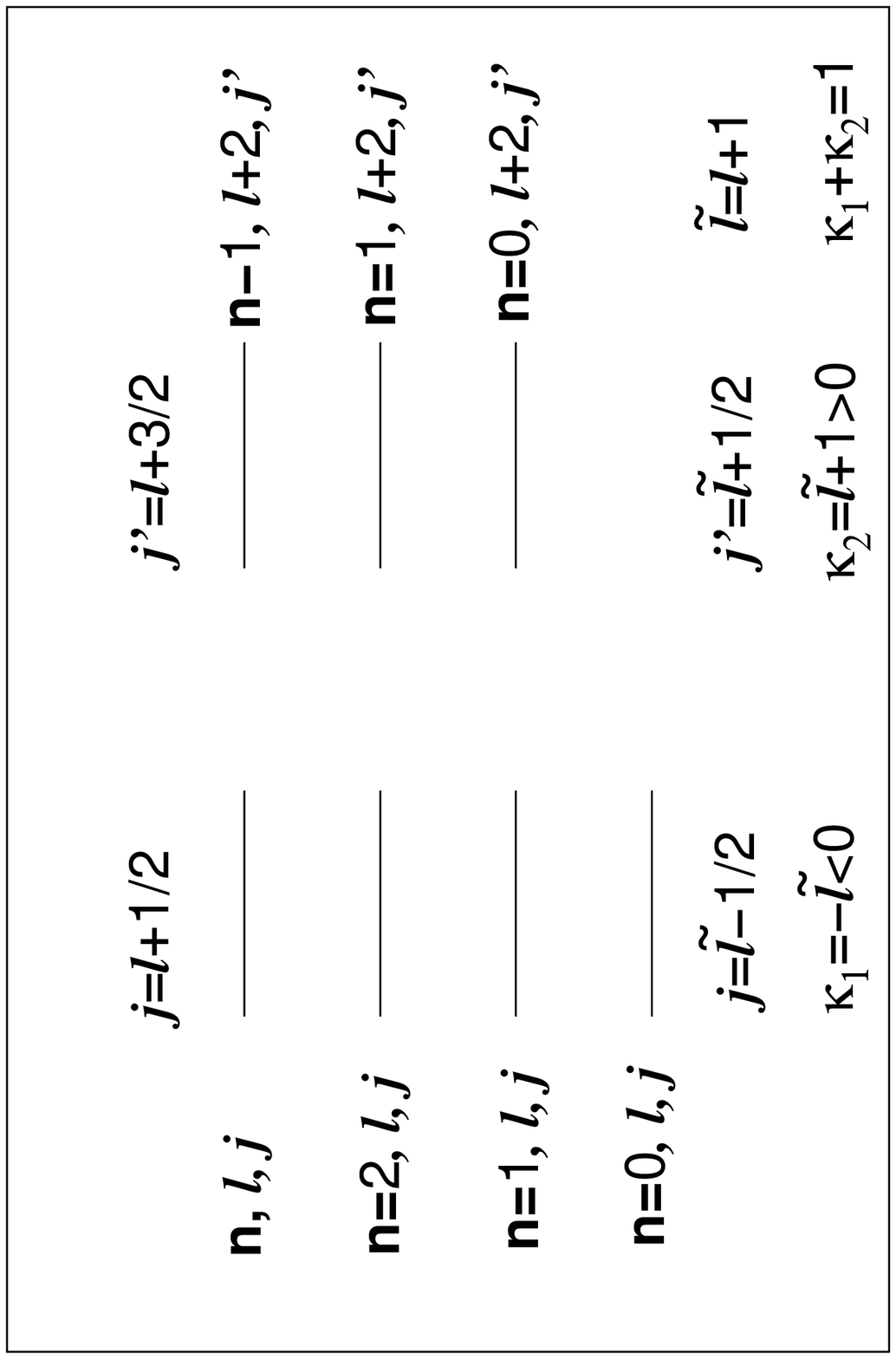}
\end{rotate}
\vspace{4cm}
\caption{Nuclear single-particle spectrum composed of pseudospin 
doublets and an ``intruder'' level. 
All states share a common $\tl$ and $\tilde{s}=1/2$. The 
corresponding Dirac $\kappa$-quantum numbers are also indicated.}
\end{minipage}
\hspace{\fill}
\hspace{0.21cm}
\begin{minipage}{0.47\linewidth}
\vspace{-0.35cm}
\epsfxsize=4.5cm
\hspace{0.15cm}
\begin{rotate}{-90}
\epsfbox{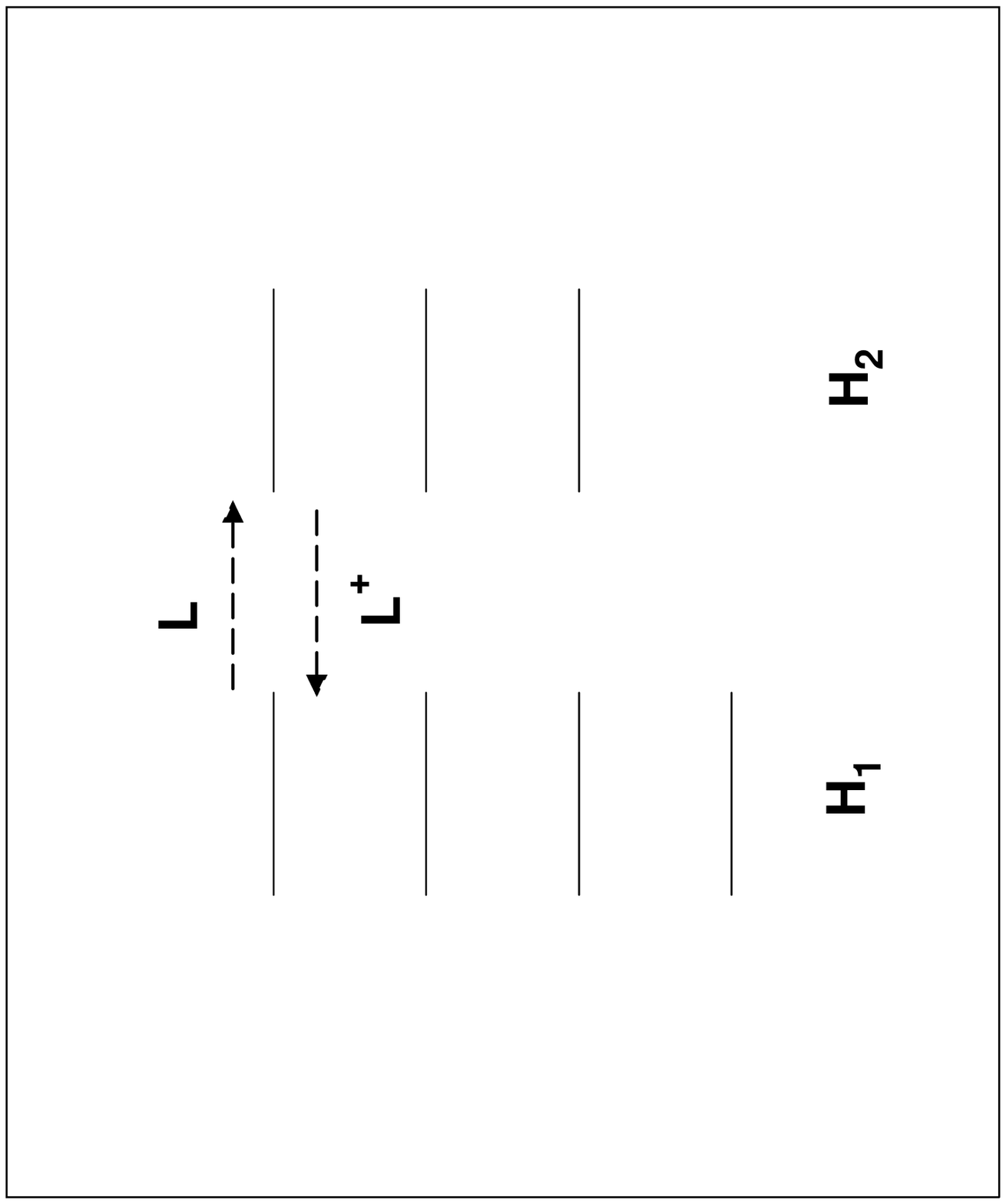}
\end{rotate}
\vspace{4cm}
\caption{Typical~supersymmetric pattern. 
The Hamiltonians $H_1$ and $H_2$ have identical spectra with an 
additional level for $H_1$ when SUSY is exact.
The operators $L$ and $L^{\dagger}$ connect the partner states.}
\end{minipage}
\hspace{\fill}
\end{figure}

\section{Supersymmetric Quantum Mechanics}
Supersymmetric quantum mechanics (SUSYQM), initially proposed as a 
model for supersymmetry (SUSY) breaking in field theory~\cite{witten81}, 
has by now developed into a field in its own right, with applications in 
diverse areas of physics~\cite{junker96}. 
The essential ingredients of 
SUSYQM are the supersymmetric charges and Hamiltonian  
\ba
Q_{-} &=& \left ({ 0\atop L}{ 0\atop 0}\right ) \;\; , \;\; 
Q_{+} = \left ({ 0\atop 0}{ L^{\dagger}\atop 0}\right ) \;\; , \;\; 
\mathcal{H} = \left ({H_1\quad\atop 0}{0\atop H_2}\right ) = 
\left ({L^{\dagger}L\atop 0}{0\atop LL^{\dagger}}\right ) \;\; 
\qquad 
\label{susyqm}
\ea
which generate the supersymmetric algebra
\ba
\left [\,\mathcal{H},Q_{\pm}\,\right ] = 
\left\{\,Q_{\pm},Q_{\pm}\,\right\}=0 \;\; , \;\; 
\left\{\,Q_{-},Q_{+}\,\right\}= \mathcal{H} ~.
\qquad
\label{susyalg}
\ea 
The partner Hamiltonians $H_1$ and $H_2$ satisfy an 
intertwining relation,  
$LH_1 = H_2L$, 
where in one-dimension the transformation operator 
$L = \frac{d}{dx} + W(x)$ 
is a first-order Darboux transformation 
expressed in terms of a superpotential $W(x)$. The 
intertwining 
relation 
ensures that 
if $\Psi_1$ is an eigenstate of $H_1$, 
then also $\Psi_2=L\Psi_1$ is 
an eigenstate of $H_2$ with the same energy, 
unless $L\Psi_1$ vanishes or produces an unphysical state ({\it e.g.} 
non-normalizable). 
Consequently, as shown in Fig.~2, the SUSY partner Hamiltonians 
$H_1$ and $H_2$ are 
isospectral in the sense that their spectra 
consist of pair-wise degenerate levels 
with a possible non-degenerate single state in one sector (when 
the supersymmetry is exact). The wave functions of the degenerate levels 
are simply related in terms of $L$ and $L^{\dagger}$. 
Such characteristic features define 
a supersymmetric pattern. In what follows we show~\cite{lev04} that a 
Dirac Hamiltonian with pseudospin symmetry obeys an intertwining 
relation and consequently gives rise to a supersymmetric pattern. 

\section{Dirac Hamiltonian with Central Fields}

A relativistic mean field description of nuclei employs a 
Dirac Hamiltonian, 
$H = \hat{\bm{\alpha}}\bm{\cdot p}
+ \hat{\beta} (M  + V_S) + V_V$, 
for a nucleon of mass~$M$ 
moving in external scalar, $V_S$, and vector,
$V_V$, potentials. 
When the potentials are spherically symmetric: $V_S=V_S(r)$, $V_V=V_V(r)$, 
the operator 
$\hat{K} = -\hat{\beta}\,(
\bm{\sigma\cdot\ell} + 1)$, 
(with $\bm{\sigma}$ the Pauli matrices and 
$\bm{\ell} = -i\bm{r}\times \bm{\nabla}$), 
commutes with $H$ and 
its non-zero integer eigenvalues $\kappa = \pm (j+1/2)$ are used to label 
the Dirac wave functions
\ba
\Psi_{\kappa,\,m} = 
\frac{1}{r} \left (
G_{\kappa} [\,Y_{\ell}\,\chi\,]^{(j)}_{m}
\atop
iF_{\kappa}[\,Y_{\ell^{\prime}}\,\chi\,]^{(j)}_{m}
\right) ~.
\label{diracwf}
\ea 
Here $G_{\kappa}(r)$ and $F_{\kappa}(r)$ 
are the radial wave functions of the upper and lower components 
respectively, $Y_{\ell}$ and $\chi$ are the spherical harmonic and 
spin function which are coupled to angular momentum $j$ with 
projection $m$.
The labels $\kappa = -(j+1/2)<0$ and $\ell^{\prime}= \ell +1$ hold 
for aligned spin 
$j=\ell+1/2$ ($s_{1/2}, p_{3/2}$, etc.), while $\kappa = (j+1/2)>0$ 
and $\ell^{\prime}= \ell -1$ hold 
for unaligned spin $j=\ell-1/2$ 
($p_{1/2},d_{3/2},$ etc.). 
Denoting the pair of radial wave functions~by
\ba
\Phi_{\kappa} = \left (
{G_{\kappa}\atop F_{\kappa}}
\right ) ~,
\label{radialwf}
\ea
the radial Dirac equations can be cast in Hamiltonian form, 
\ba
H_{\kappa}\,\Phi_{\kappa} &=& E\,\Phi_{\kappa} ~,
\label{hradial}
\ea
with
\bmath
\ba
H_{\kappa} &=& 
\left (
\begin{array}{ll}
M + \Delta   & \quad -\frac{d}{dr} + \frac{\kappa}{r} \\
\frac{d}{dr} + \frac{\kappa}{r} & \quad -(M + \Sigma) 
\end{array}
\right )
\label{hk}\\
\Delta(r) &=& V_S + V_V\;\ , \;\; \Sigma(r) = V_S - V_V.
\label{delsig}
\ea
\label{hkgen}
\emath
The nuclear single-particle spectrum is obtained from 
the valence bound-state solutions of Eq.~(\ref{hradial}) 
with positive 
binding energy $(M-E) > 0 $ and total energy $E>0$. 
The non-relativistic shell-model wave functions 
are identified with the upper components of the 
Dirac wave functions (\ref{diracwf}). 
For relativistic mean fields relevant to nuclei, 
$V_S$ is attractive and $V_V$ is repulsive with typical values 
$V_S(0) \sim -400,\; V_V(0)\sim 350,$ MeV. 
The potentials satisfy 
$rV_S,\; rV_V \rightarrow 0$ for $r\rightarrow 0$, 
and $V_S,\; V_V \rightarrow 0$ for $r\rightarrow \infty$. 
Under such circumstances 
one can prove~\cite{levgino01} the following properties which 
are relevant for understanding the nodal structure of 
pseudospin doublets and intruder levels in nuclei.
\begin{alphlist}[(b)]
\item The radial nodes of $F_{\kappa}$ ($n_F$) and 
$G_{\kappa}$ ($n_G$) are related. Specifically,  
\ba
\begin{array}{ll}
n_F = n_G \qquad & {\rm for}\;\;\kappa < 0  ~,\\ 
n_F = n_G + 1 \qquad & {\rm for}\;\; \kappa > 0 ~. \\
\end{array}
\label{nodes}
\ea
\item Bound states with $n_F=n_G=0$ can occur only for $\kappa<0$.
\end{alphlist}

\section{Relativistic Pseudospin Symmetry in Nuclei}
A relativistic pseudospin symmetry occurs when 
the sum of the scalar and vector potentials is a constant 
\ba
\Delta(r) = V_{S}(r) + V_{V}(r) = \Delta_0 ~.
\label{pspot}
\ea 
A Dirac Hamiltonian satisfying (\ref{pspot}) has an invariant 
$SU(2)$ algebra 
generated by~\cite{ginolev98} 
\ba
{\hat{\tilde {S}}}_{\mu} =
\left (
\begin{array}{cc}
\hat {\tilde s}_{\mu} &  0 \\
0 & {\hat s}_{\mu}
\end{array}
\right ) ~.
\label{Sgen}
\ea
Here ${\hat s}_{\mu} = \sigma_{\mu}/2$ are the usual spin operators, 
$\hat {\tilde s}_{\mu}= U_p {\hat s}_{\mu} U_p$ 
and $U_p = \frac{\bm{\sigma\cdot p}}{p}$ is a
momentum-helicity unitary operator~\cite{draayer}.
In the symmetry limit the Dirac eigenfunctions belong to the spinor 
representation of $SU(2)$. 
The relativistic pseudospin symmetry determines the form of the 
eigenfunctions in the doublet to be 
\bmath
\ba
\Psi_{\kappa_1<0,\,m} = 
\frac{1}{r} \left (
G_{\kappa_1} [\,Y_{\tl-1}\,\chi\,]^{(j)}_{m} 
\atop
iF_{\kappa_1}[\,Y_{\tl}\,\chi\,]^{(j)}_{m}
\right) \;\;\; \kappa_1 &=& -\tl<0\;,\; j=\tl-1/2 ~,
\label{wfnegk}
\\[5pt]
\Psi_{\kappa_2>0,\,m} =
\frac{1}{r} \left (
G_{\kappa_2} [\,Y_{\tl+1}\,\chi\,]^{(j^{\prime})}_{m}
\atop
iF_{\kappa_2}[\,Y_{\tl}\,\chi\,]^{(j^{\prime})}_{m}
\right) \;\;\; \kappa_2&=& \tl+1>0\;,\; j^{\prime}=\tl+1/2 ~,
\qquad\qquad 
\label{wfposk}
\ea 
\label{wfnegposk}
\emath
\begin{figure}[t]
\begin{minipage}{0.95\linewidth}
\epsfxsize=12cm
\epsfbox{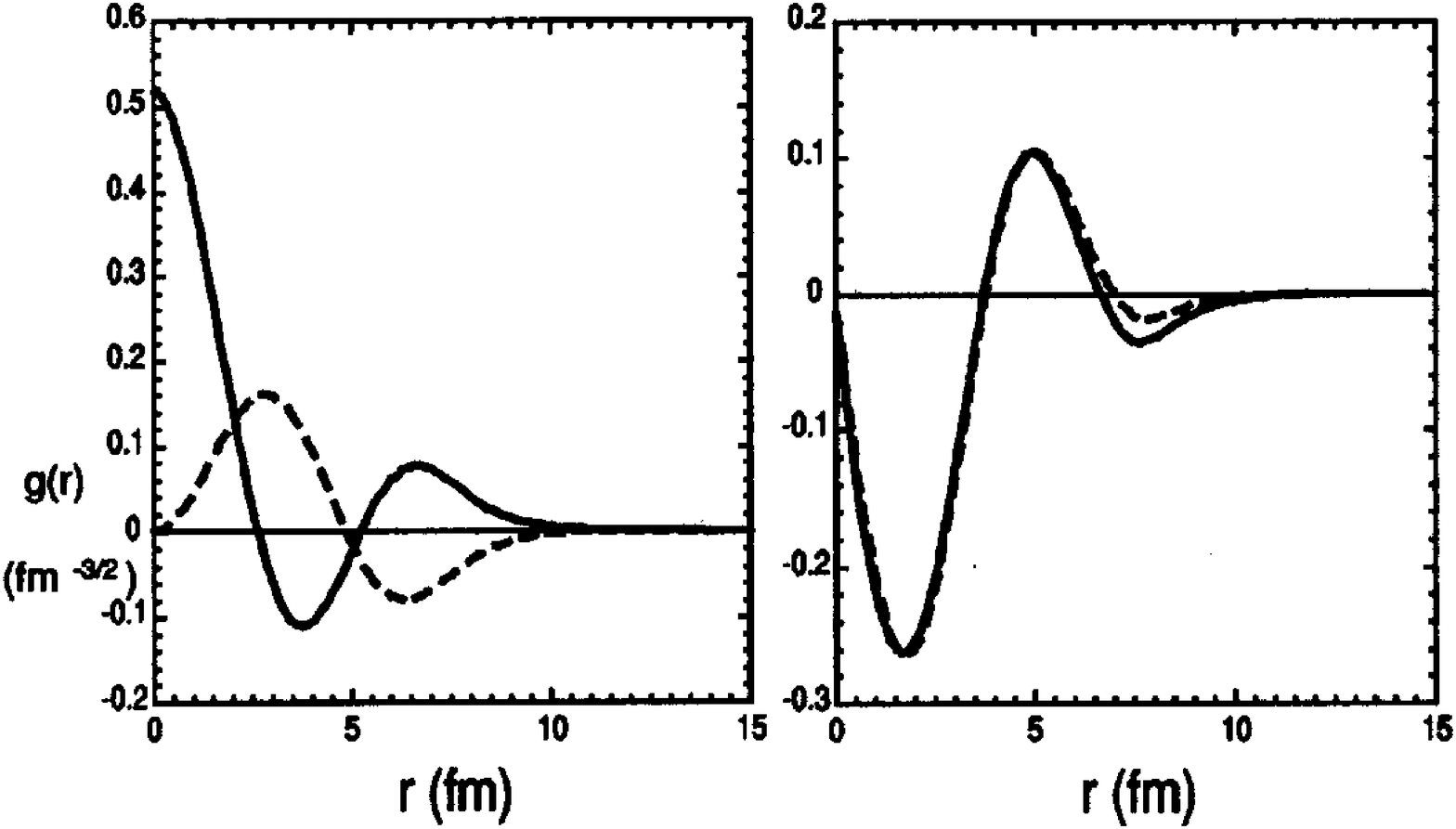}
\end{minipage}
\hspace{\fill}
\begin{minipage}{0.48\linewidth}
\epsfxsize=6.2cm
\epsfbox{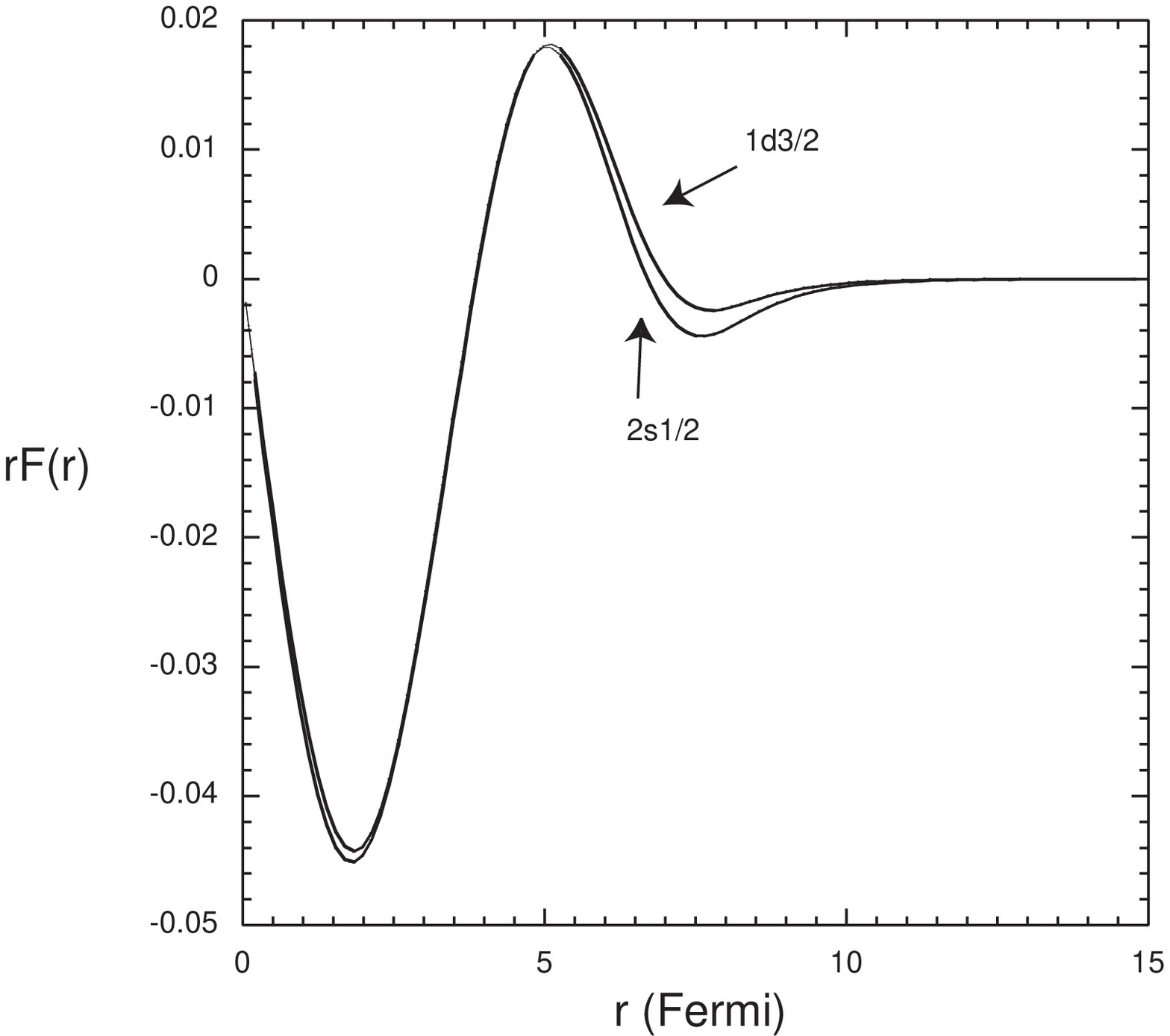}
\end{minipage}
\hspace{\fill}
\hspace{0.4cm}
\begin{minipage}{0.4\linewidth}
\vspace{-0.6cm}
\caption{Top left panel: the upper components 
$g(r)=rG_{\kappa}(r)$ of the $2s_{1/2}$, (solid line) and $1d_{3/2}$ 
(dashed line) Dirac eigenfunctions in $^{208}$Pb. 
Top~right panel: testing 
the differential relation of Eq.~(\ref{gcond}) 
for the upper components of $2s_{1/2}$ ($\kappa_1=-1$) 
and $1d_{3/2}$ ($\kappa_2=2$). Bottom panel: the lower components 
$f(r)=rF_{\kappa}(r)$ of $2s_{1/2}$ and 
$1d_{3/2}$, testing relation (\ref{fcond}). 
Based on calculations in [12,13].}
\end{minipage}
\hspace{\fill}
\end{figure}
and imposes the following conditions on their radial amplitudes:  
\bmath
\ba
F_{\kappa_1} &=& F_{\kappa_2} ~,
\label{fcond}
\\
\frac{dG_{\kappa_1}}{dr} + \frac{\kappa_1}{r}G_{\kappa_1} 
&=& \frac{dG_{\kappa_2}}{dr} + \frac{\kappa_2}{r}G_{\kappa_2} ~.
\label{gcond}
\ea
\label{fgcond}
\emath
The two 
eigenstates in the doublet are connected by the pseudospin generators 
${\hat{\tilde {S}}}_{\mu}$ (\ref{Sgen}). 
The lower components are connected by the usual spin operators and, 
therefore, have the same spatial wave functions. 
Consequently, the two states of the doublet share a common $\tl$ which is 
the orbital angular momentum of the lower component. The Dirac structure 
then ensures that the orbital angular momentum of 
the upper components in Eq.~(\ref{wfnegposk}) 
is $\ell= \tl-1$ for $j=\tl-1/2=\ell+1/2$, and 
$\ell+2=\tl+1$ for $j^{\prime}=\tl +1/2=\ell+ 3/2$. 
This explains the particular angular momenta defining the 
pseudospin doublets in Eq.~(\ref{psdoub}). 
The radial amplitudes of the lower components are equal (\ref{fcond}) 
and, in particular, have the same number of nodes $n_F=n$. 
Property~(a) of the previous section then ensures that 
$G_{\kappa_1}$ has $n$ nodes and $G_{\kappa_2}$ has $n-1$ 
nodes, in agreement with Eq.~(\ref{psdoub}). 
Property~(b) ensures that the 
Dirac state with $n_F=n_G=0$, corresponding to the ``intruder'' shell-model 
state, has a wave function as in Eq.~(\ref{wfnegk}) with $\kappa<0$, 
and does not have a partner eigenstate (with $\kappa>0$). 

Realistic mean fields in nuclei approximately satisfy condition (\ref{pspot}) 
with $\Delta_0\approx 0$. The required breaking of pseudospin symmetry in 
nuclei is small. Quasi-degenerate doublets of normal-parity states and 
abnormal-parity levels without a partner eigenstate persist in the spectra. 
The relations (\ref{fgcond}) between wave functions 
have been tested in numerous realistic mean field calculations in a 
variety of nuclei, and were found to be obeyed to a good approximation, 
especially for doublets near the Fermi surface~\cite{ginomad98,ginolev01}. 
A representative example for neutrons in $^{208}$Pb is shown in Fig.~3. 

\section{Relativistic Pseudospin Symmetry and SUSY}
In the pseudospin limit, Eq.~({\ref{pspot}), the two Dirac states 
$\Psi_{\kappa_1<0,m}$ and $\Psi_{\kappa_2>0,m}$ of Eq.~(\ref{wfnegposk}) 
with $\kappa_1+\kappa_2=1$ are degenerate, unless both the upper and 
lower components have no nodes, in which case only $\Psi_{\kappa_1<0,m}$ 
is a bound state. 
Altogether, as shown in Fig.~4, the ensemble of Dirac 
states with $\kappa_1+\kappa_2=1$ 
exhibits a supersymmetric pattern of twin towers with pair-wise degenerate 
pseudospin doublets sharing a common $\tl$, and an additional 
non-degenerate nodeless state at the bottom of the $\kappa_1<0$ tower. 
An exception to this rule 
is the tower with $\kappa_2=1$ ($p_{1/2}$ states with $\tl=0$), 
which is on its own, because states with $\kappa_1=0$ do not exist. 
The supersymmetric structure arises because in the pseudospin limit 
the respective radial Dirac Hamiltonians, $H_{\kappa_1}$ and 
$H_{\kappa_2}$, satisfy an intertwining relation of the form 
\ba
LH_{\kappa_1} = H_{\kappa_2}L ~,
\label{lh1h2}
\ea
with $\kappa_1+\kappa_2=1$.
\begin{figure}[t]
\epsfxsize=6cm
\hspace{2cm}
\begin{rotate}{-90}
\epsfbox{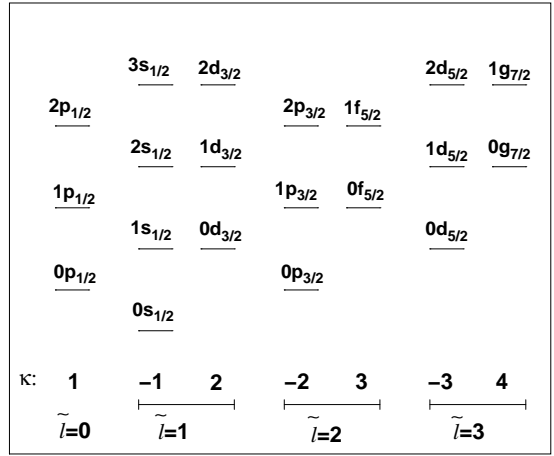}
\end{rotate}
\vspace{6.0truecm}
\caption{Schematic supersymmetric pattern in the pseudospin limit 
of the Dirac Hamiltonian [8]. \label{Fig4}} 
\end{figure}
The transformation operator is found to be
\ba
L = b \left (
\begin{array}{cc}
0 \quad & \frac{d}{dr} - \frac{\kappa_2}{r} \\
-\frac{d}{dr} - \frac{\kappa_1}{r} \quad & 
(2M + \Sigma + \Delta_0)
\end{array}
\right ) ~.
\label{Ldel0}
\ea
$L$ connects the two doublet states 
\ba
L\,\Phi_{\kappa_1} = b(M + \Delta_0 - E)\,\Phi_{\kappa_2} 
\qquad (\kappa_1+\kappa_2=1) ~,
\label{LGF}
\ea
and identifies the two states as 
supersymmetric partners. 
Eq.~(\ref{LGF}) relies on the input that the two states are 
eigenstates of the Dirac Hamiltonian with $E_{\kappa_1}=E_{\kappa_2}=E$ 
and their wave functions satisfy the 
relations in Eq.~(\ref{fgcond}). The fact that nuclear wave 
functions, obtained in realistic mean-field calculations, obey 
these relations to a good approximation, confirms the 
relevance of supersymmetry to these nuclear states.

Constructing supersymmetric charges $Q_{\pm}$ and Hamiltonian 
$\mathcal{H}$ from $L$ and $H_{\kappa_1}$, $H_{\kappa_2}$ as in 
Eq.~(\ref{susyqm}), ensures the fulfillment of Eq.~(\ref{susyalg}), 
except for the last relation which now reads 
\ba
\{Q_{-},Q_{+}\} = 
b^2 [\mathcal{H} - (M+\Delta_0)][\mathcal{H}- (M+\Delta_0)] ~. 
\label{susydef}
\ea
Eq.~(\ref{susydef}) involves a polynomial of $\mathcal{H}$, 
indicating a quadratic deformation 
of the conventional supersymmetric algebra~\cite{deber02}. The latter 
arises because both the Dirac Hamiltonian, $H_{\kappa}$, and the 
transformation operator, $L$, are of first order. 
In real nuclei, the relativistic pseudospin symmetry is slightly broken, 
implying $\Delta(r)\neq \Delta_0$ in Eq.~(\ref{pspot}). 
Taking $H_{\kappa}$ as in Eq.~(\ref{hkgen}) and 
$L$ as in Eq.~(\ref{Ldel0}) but with 
$\Delta_0\rightarrow\Delta(r)$, we now find that 
\ba
LH_{\kappa_1} - H_{\kappa_2}L = i\,b\,
\frac{d\Delta}{dr}\,\sigma_2
\ea
Furthermore, $\{Q_{-},Q_{+}\}$ has the same formal form as in 
Eq.~(\ref{susydef}), but the appearance of $\Delta(r)$ instead of 
$\Delta_0$ implies that the anticommutator is no longer just a 
polynomial of $\mathcal{H}$. 

\section{Summary}
We discussed a possible grouping of shell-model single-particle 
states into larger multiplets, exhibiting a supersymmetric pattern. 
The multiplets involve several quasi-degenerate pseudospin doublets 
and intruder levels without a partner eigenstate.
In contrast to previous studies of pseudospin in nuclei, the suggested 
grouping of nuclear states treats the intruder levels and pseudospin
doublets on equal footing. 
The underlying supersymmetric structure is linked with an approximate 
relativistic pseudospin symmetry of the nuclear mean-field Dirac Hamiltonian. 
The relativistic pseudospin symmetry imposes relations between the upper 
and lower components of the two Dirac states forming the doublet. 
These relations, which are obeyed to a good approximation by realistic 
mean-field wave functions, 
imply that the Dirac Hamiltonian with pseudospin symmetry obeys 
an intertwining relation which gives rise to the indicated 
supersymmetric pattern.

\section*{Acknowledgments}
This work was partly supported by the U.S.-Israel Binational Science 
Foundation, in collaboration with J.N. Ginocchio (LANL), and partly  
by the Israel Science Foundation.

\end{document}